\documentstyle[11pt,moriond,epsfig]{article}

\def\Journal#1#2#3#4{{#1} {\bf #2}, #3 (#4)}

\def\NIMA{{\em Nucl. Instrum. Methods} A}

\def\PLB{{\em Phys. Lett.}  B}
\def\PRL{\em Phys. Rev. Lett.}

\def\PRB{{\em Phys. Rev.} B}

\def\APJS{\em Astrophys. J. Sup.}
\def\AA{\em A\&A}

\def\be{\begin{equation}}
\def\ee{\end{equation}}
\def\bea{\begin{eqnarray}}
\def\eea{\end{eqnarray}}

\newcommand{\Au} {\mbox{$ ^{197}{\rm{Au}}$}~}

\newcommand{\neut}{$\tilde{\chi}$~}
\newcommand{\neu}{$\tilde{\chi}$}
\newcommand{\Co} {\mbox{$ ^{57}{\rm{Co}}$}~}

\newcommand{\hetrois}    {\mbox{$ ^{3}{\mathrm{He}}
$}~}

\newcommand{\gam}{\mbox{\rm $\gamma$-ray}~}
\newcommand{\gams}{\mbox{\rm $\gamma$-rays}~}

\newcommand{\microns}{\mbox{{\rm$\mu$m}}~}
\newcommand{\muK}{\mbox{{\rm $\mu$K}}~}
\newcommand{\muKs}{\mbox{{\rm $\mu$K}}}
\begin{document}
\vspace*{4cm}
\title{MACHE3, \\A PROTOTYPE FOR NON-BARYONIC DARK MATTER SEARCH:\\
KeV EVENT DETECTION AND MULTICELL CORRELATION}

\author{ C. Winkelmann$^1$, E. Moulin$^2$,}
\author{Yu. Bunkov$^1$, J. Genevey$^2$, H. Godfrin$^1$, J. 
Mac\'{\i}as-P\'erez$^2$, J.A. Pinston$^2$, D. Santos$^2$}

\address{$^1$ Centre de Recherches sur les Tr\`es Basses Temp\'eratures,
  CNRS and Universit\'e Joseph Fourier, BP166, 38042 Grenoble cedex 9, 
France \\$^2$ Laboratoire de Physique Subatomique et de Cosmologie,
  CNRS/IN2P3 and Universit\'e Joseph Fourier,
  53, avenue des Martyrs, 38026 Grenoble cedex, France\\}

\maketitle\abstracts{
%Superfluid \hetrois at ultra-low temperatures (100 \muKs)
%is a sensitive medium for the bolometric detection of particles.
MACHe3 (MAtrix of Cells of Helium 3) is a project
for non-baryonic dark matter search using \hetrois as a sensitive medium.
Simulations made on a high granularity detector show a very good rejection
to background signals.
A multicell prototype including 3 bolometers has been developed
to allow correlations between the cells for background event discrimination.
One of the cells contains a low activity \Co source providing conversion
electrons of 7.3 and 13.6 keV to confirm the detection of low energy events.
First results on the multicell prototype are presented. A
detection threshold of 1 keV has been achieved.
The detection of low energy conversion electrons coming from the
\Co source is highlighted as well as the cosmic muon spectrum
measurement. The
possibility to reject background events  by using the correlation
among the cells
is demonstrated from the simultaneous detection of muons in different cells.}
\section{Introduction}
 From the latest results from observational cosmology experiments,
the determination of the cosmological parameters has reached an 
unprecedented level
of accuracy. Measurements of the cosmic microwave background (CMB) anisotropies
~\cite{wmap,archeops} used in combination with
large scale structures surveys
and supernovae measurements point out that most of the matter is composed of
non-baryonic dark matter.
The leading candidates intended to explain this yet undiscovered
component consist of
new particles referred to as WIMPs (Weakly Interacting Massive Particles).
Supersymmetric extensions of the standard model of particle physics provide
a compelling dark matter candidate~\cite{jungman},
the lightest neutralino \neut, a
neutral and colorless linear
superposition of the superpartners of the gauge (B, {\rm W$^3$}) and
Higgs ({\rm H$^0_d$,
H$^0_u$})
bosons.

Many collaborations have developed promising detectors to search for
non-baryonic dark matter.
These detectors have reached sufficient sensitivity to begin to test
regions of the SUSY parameter
space. Direct detection experiments present common problems such as
the neutron interaction
background and the radioactivity contamination from both the
sensitive medium and the surrounding materials.
Based on early experimental works, a superfluid \hetrois detector
for direct detection of non-baryonic dark matter has been 
proposed~\cite{lanc,Second}. First experimental tests of a \hetrois 
detector by neutrons and $\gamma$-rays have been done in Lancaster and 
Grenoble~\cite{lanc1,nature95,PRB98}.
Monte Carlo simulations have demonstrated that a high granularity
\hetrois detector
would allow to reach high rejection factors against background events
by using the measurement of the released energy together with the
correlation among the cells~\cite{nima}.
Such a technique permits to obtain a low false neutralino event rate.

At ultra-low temperatures, \hetrois is a very appealing material
because it constitutes a highly sensitive bolometer.
At temperatures around 100 $\mathrm{\mu}$K, \hetrois is in its
superfluid B phase
for which the quasiparticle gap and the heat capacity are extremely small.
The use of \hetrois
profits of very interesting features with respect to other materials~:\\
\indent- \hetrois being a 1/2 spin nucleus, a \hetrois detector will
be mainly sensitive
to the axial interaction, making this device complementary to
existing ones, mainly sensitive to the scalar
interaction. The axial
interaction is largely dominant in all the SUSY region associated
with a substantial elastic cross-section~\cite{plbf}.\\
\indent- A close to absolute purity (nothing can dissolve in \hetrois
at 100 $\mathrm{\mu}$K).\\
\indent- A high neutron capture cross-section,
leading to a large energy release through the nuclear reaction {\rm
n\,+\,\hetrois\,$\rightarrow$\,p\,+$^3$H\,+\,764\,keV}.
Neutron contamination has thus a clear signature~\cite{nima,nature95}, well
discriminated from a WIMP
signal.  \\
\indent- Low Compton cross-section
. No intrinsic X-rays.\\
\indent- A high signal to noise ratio, due to the narrow energy range
expected for a WIMP signal.
The maximum recoil energy does only slightly depend on the
WIMP mass, due to the fact that the target nucleus (${\rm m=2.81 \,
GeV}\!/c^2$) is much
lighter than the incoming  \neu. As a matter of fact, the recoil
energy range needs~\cite{nima,plbf} to be studied only below ${\rm 6 \,keV}$.

\section{Bolometric particle detection with superfluid $^3$He}
\subsection{Vibrating Wire Thermometry}
\label{thermometry}
\begin{figure}[t]
\centerline{\includegraphics[width=0.45\textwidth]{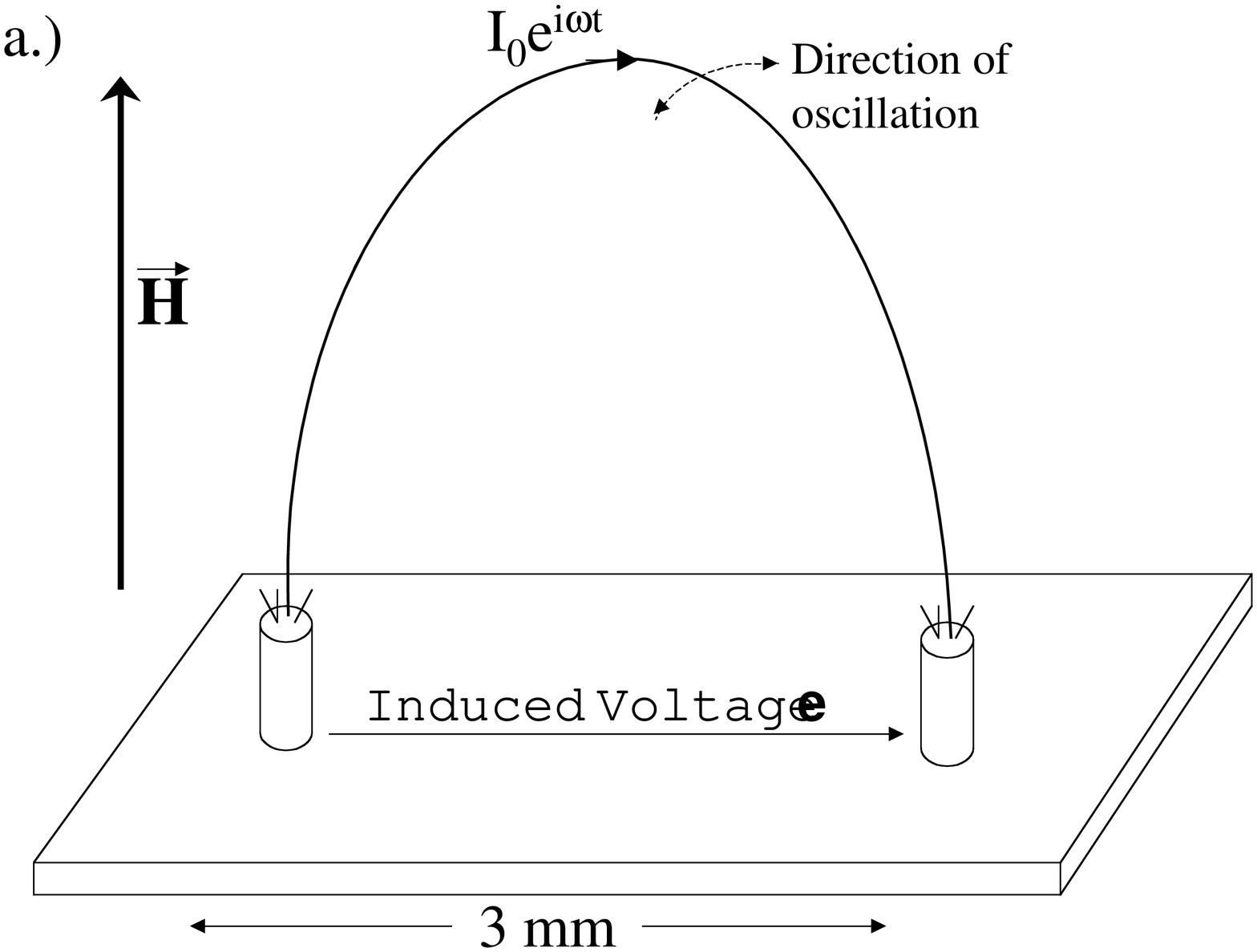}
\hspace{1cm}
\includegraphics[width=0.45\textwidth]{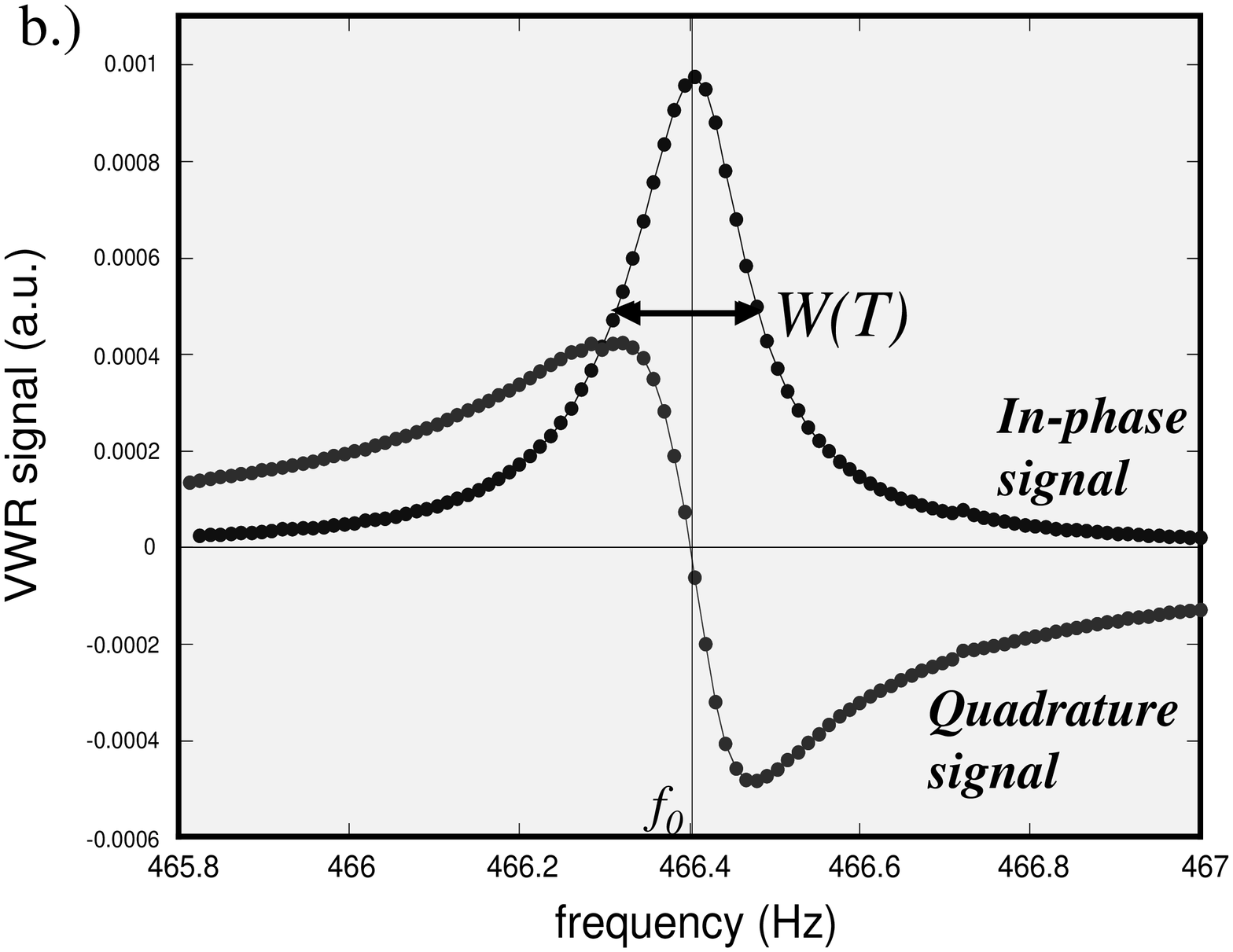}}
\caption{Setup for a monofilamentary NbTi Vibrating Wire (a). The
induced signal is proportional to the oscillation amplitude and
inversely proportional to the damping.
The plot of a typical
resonance sweep (b) with $W(T)$=170 mHz indicates here a temperature
of about 120 $\mu$K.}
\label{VWR}
\end{figure}

A Vibrating Wire Resonator (VWR), sketched on the figure~\ref{VWR}a, is a
fine superconducting
   wire bent into semi-circular shape and oscillating perpendicularly
to its plane.
   The oscillations are excited by a Laplace force in a uniform
external magnetic field of
   about 100 mT, similarly to a loud-speaker. Typical resonant
frequencies for the 4.5 \microns
   diameter NbTi VWRs used are of about 500 Hz. The amplitude of the
oscillations are detected
   via the voltage induced by the motion of the wire through the field
lines. The signal is
   amplified by a cold transformer and a room-temperature low-noise
pre-amplifier before being
   read out by a lock-in amplifier.

The dynamics of the VWR can be conveniently described by a damped
harmonic oscillator model.
   The damping of the VWR is due to friction with the quasiparticles
(QPs) of the surrounding
   superfluid~\cite{lancaster}. Due to the presence of an isotropic
superfluid gap of order $\Delta=0.14\ \mu eV$ in $^3$He-B, the QP
density and thus the friction decrease exponentially as temperature
is lowered. The resonance line-width at half-height $W$
(figure~\ref{VWR}b) is proportional to the QP-damping and related to
temperature by $W(T)=W_0 \exp(   -\Delta/k_BT).$
It exhibits an extremely steep temperature dependence around 100 $\mu$K.
The measurement of $W$ allows to access temperature with response
times $\tau_{wire}=1/\pi W < 1$ s.

\subsection{$^3$He bolometry}
We use copper boxes of typical dimensions about 5 mm, filled with
$^3$He which is in weak
thermal contact with the outer bath through a small orifice. The
interaction of a particle with
the $^3$He in the box releases energy which results in an increase of
the QP temperature, and
thus density. The time constants for internal equilibrium of the
quasi-particle gas are small
($<$ 1 ms).
The time constant for thermal relaxation of QPs through the orifice after a
heating event was
tuned to be $\tau_{box} \approx 5$ s. The heat leak through the
container walls can be neglected
   because of the huge thermal resistance (Kapitza resistance) of
solid-liquid interfaces at very
   low temperatures. Each box contains a least one VWR-thermometer
which allows to follow rapid
   variations of the temperature.

\begin{figure}
\begin{center}
\psfig{figure=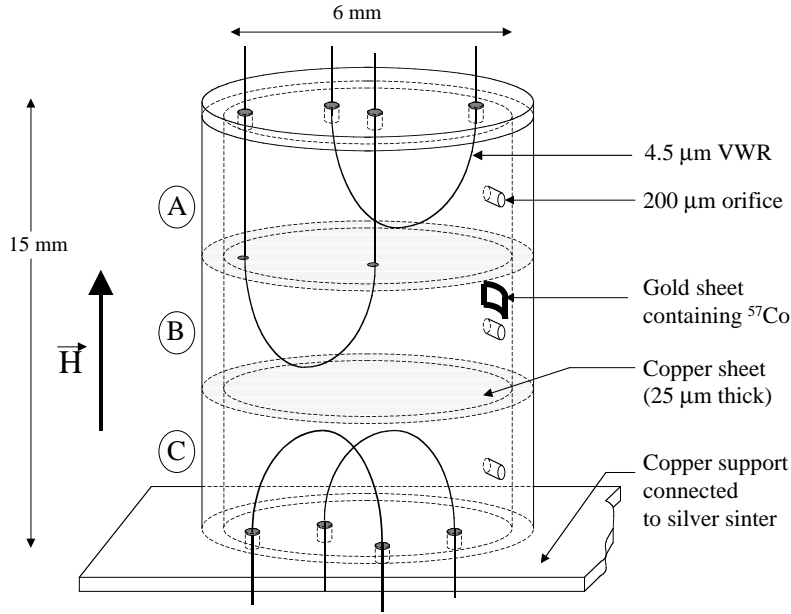,height=3.3in}
\caption{Experimental setup of the 3-cell bolometer. The 3 adjacent
copper boxes are filled with superfluid \hetrois in weak thermal
contact through the orifices with the outer bath, which acts as a
thermal reservoir. Cell 'B' contains a low activity
$^{57}$Co source (0.1 Bq). Cell 'C' contains an extra VWR for calibration.}
\label{fig:3cell}
\end{center}
\end{figure}

In order to demonstrate the possibility of efficient background
rejection by correlating many cells,
   we have built a 3-cell bolometer prototype, as sketched in the
   figure~\ref{fig:3cell}. One of the cells (C)
   contains an extra VWR, slightly larger, that allows to calibrate the
bolometer by depositing a well
   controlled amount of heat during a short mechanical pulse.
One of the key features of this setup is the presence of a low activity
\Co source
in the cell B. By detecting the low energy electrons
emitted by this source we show that we have achieved a detection
threshold of the order of 1 keV.
\section{Experimental results}
\subsection{Analysis method}
\begin{figure}[h]
\begin{center}
\psfig{figure=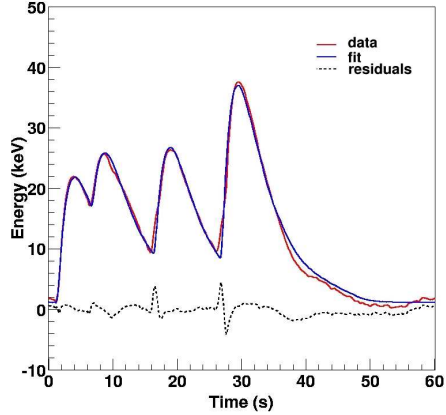,height=2.25in}
\caption{Example of an analysis of a data sample at 100 \muK
coming from the cell containing the \Co source.
The spectrum presents low energy events with an important pile-up.
Denoised data are in red and the fit in blue. Dotted line corresponds
to fitting residuals.
Detected peaks have a signal to noise ratio higher than 5.\label{fig:fit}}
\end{center}
\end{figure}

The raw data (see figure~\ref{fig:fit}) consists of a serie of peaks
which corresponds
to the energy released by particles interacting with the \hetrois
inside the cell.
A procedure has been developed to analyse the spectra. Important efforts
have been done to perform a correct treatment of low energy events
presenting a high rate of pile-up.
The method is based on several steps. First, a wavelet denoising allows to
reduce significantly the noise on the raw data.
The baseline, corresponding to
low frequency temperature fluctuations in the thermal bath,
is then removed. An iterative
fit including a peak flagging followed by a fit to a reference peak
previously extracted
from data, allows to retrieve the position, amplitude and signal to noise
ratio (S/N) for each detected peak~\cite{moulin}.

\subsection{Low energy spectrum from the low activity \Co source}
\begin{figure}[!ht]
\begin{center}
\psfig{figure=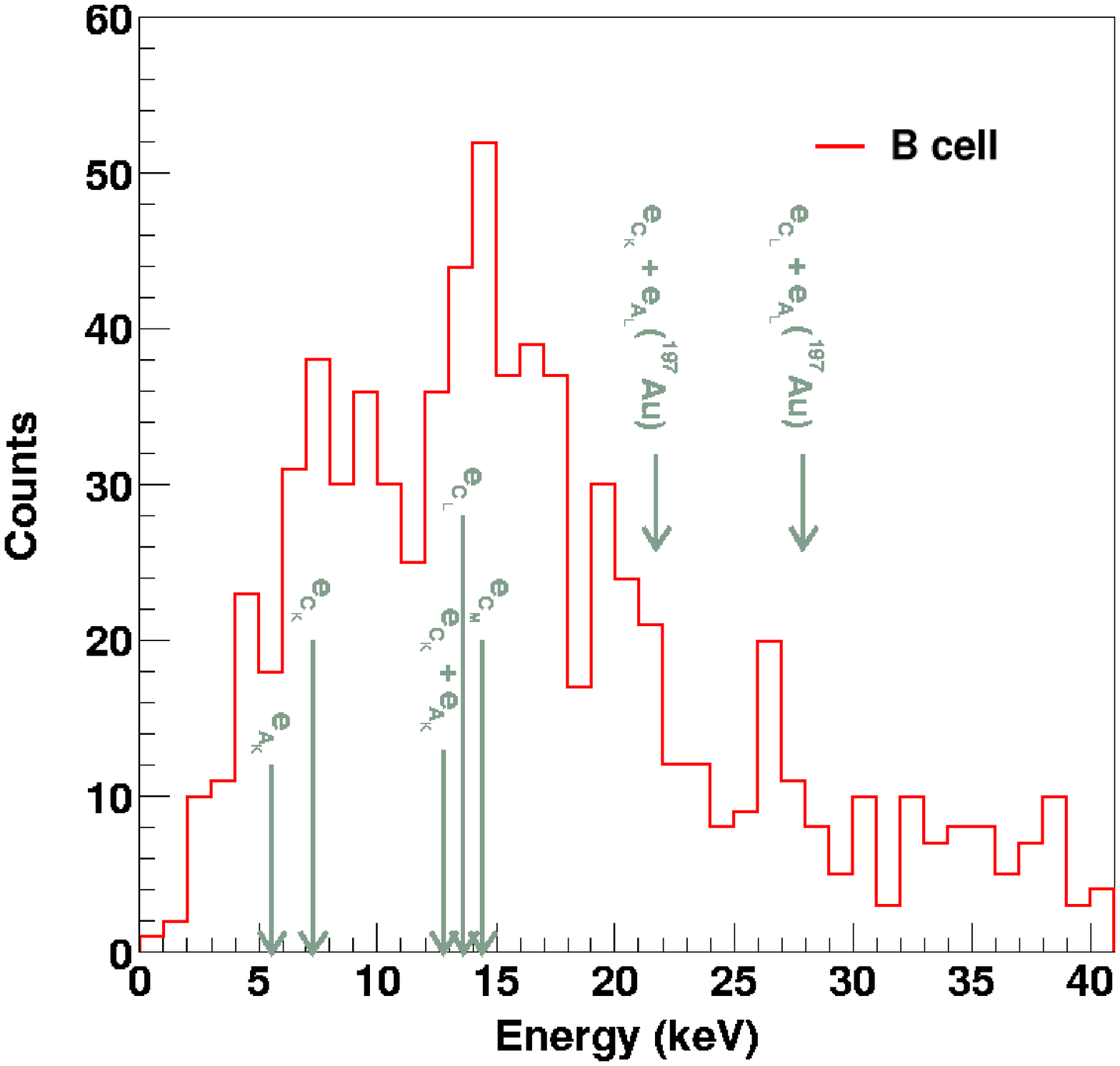, height=2.5in}
\psfig{figure=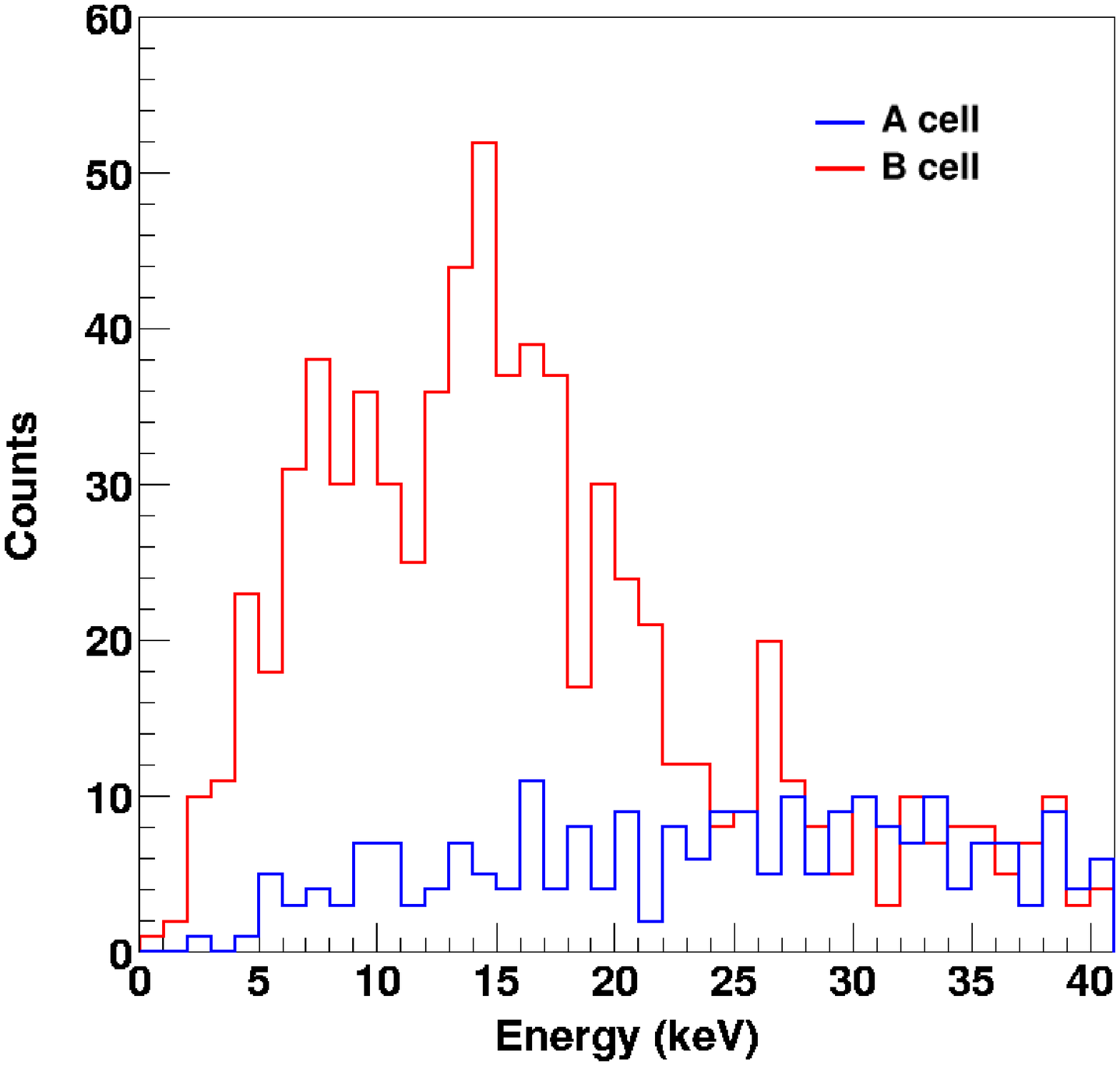, height=2.5in}
\caption{Low energy spectrum in the cell containing the \Co source (red line).
Two main structures are clearly visible around 7.3 and 13.6 keV corresponding
to the conversion electrons from the K and L shell of 14 keV nuclear transition
respectively. Two
other structures are
seen around 20 and 27 keV coming from the pile-up between the L shell
Auger electron
from
\Au and the K and L shell conversion electrons respectively.
Spectrum in the cell without source is shown on the right plot (blue line).
This corresponds to background events mainly composed of cosmic
muons.\label{fig:specelec}}
\end{center}
\end{figure}

In order to demonstrate the feasibility of detecting low energy events,
a very low activity internal conversion electron \Co source made in LPSC,
deposited on a thin gold foil, has been embedded in one of the
cells of the prototype as shown on figure~\ref{fig:3cell}. This low
activity source
({\rm $\sim$0.1 Bq}) provides electrons in the desired energy range
\rm{($\sim$ 6 keV)}.
Data  at 100~\muK have been taken during 18 hours. The dedicated analysis
method allowed to obtain the low energy electron
spectrum presented in figure~\ref{fig:specelec}.
The low energy electron lines are clearly  visible. The fact to see
these confirms the very low background
coming from the Compton interaction of the \gams (121 and 136 keV) of
the source.
The main spectrum structures have been identified from the nuclear
desintegration scheme of the \Co source.
The main expected lines are reminded on figure~\ref{fig:specelec}.
The spectrum  is composed of
two main components which correspond to the K shell
and L shell internal conversion electrons of 14 keV nuclear 
transition respectively at 7.3 and
13.6 keV.
Another contribution comes from the Auger electrons corresponding to the K
and L shell. Different pile-ups are expected from the desintegration scheme.
A line at 12.8 keV comes from
the superposition of the K shell conversion electron with its
corresponding 5.5 keV
Auger electron. Two more peaks are observed at 21.7 and 27.9 keV. They result
more likely from the 122 keV \gam interaction in the gold foil. 
Indeed, the L shell Auger
electron from \Au can pile-up with the K and L shell conversion
electrons respectively.
The background has
been measured in the energy range of interest. It is shown on the
right hand side
plot on
the figure~\ref{fig:specelec}. It is mainly composed of cosmic muons
crossing the cell on
peripheral tracks. Only two events on seven remain after
multicell correlation between 1 and 6 keV.
For further details, see E.~Moulin {\it et al.}~\cite{moulin}.

\subsection{Cosmic muons}
\begin{figure}[h]
\begin{center}
\psfig{figure=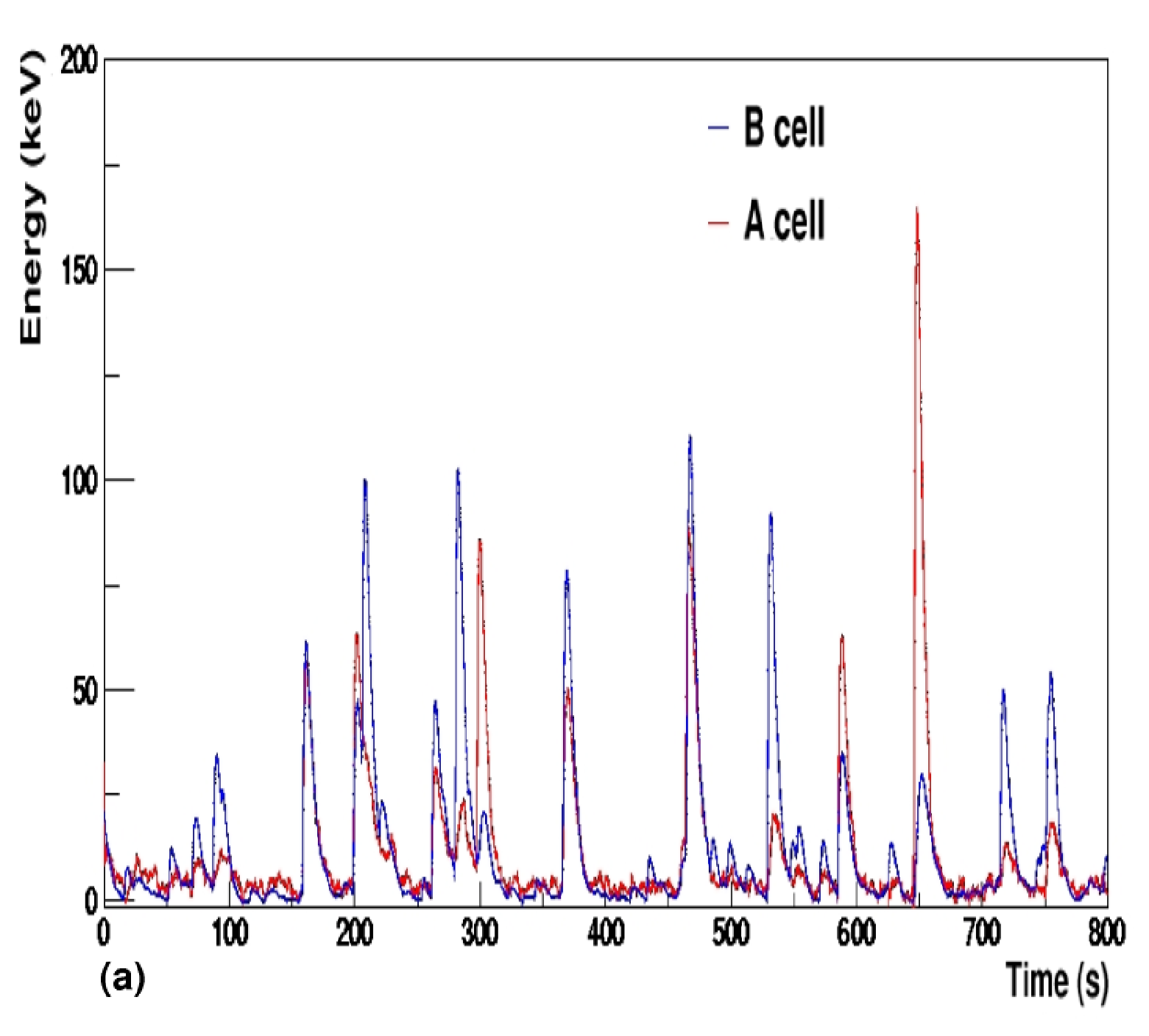,height=2.5in}
\psfig{figure=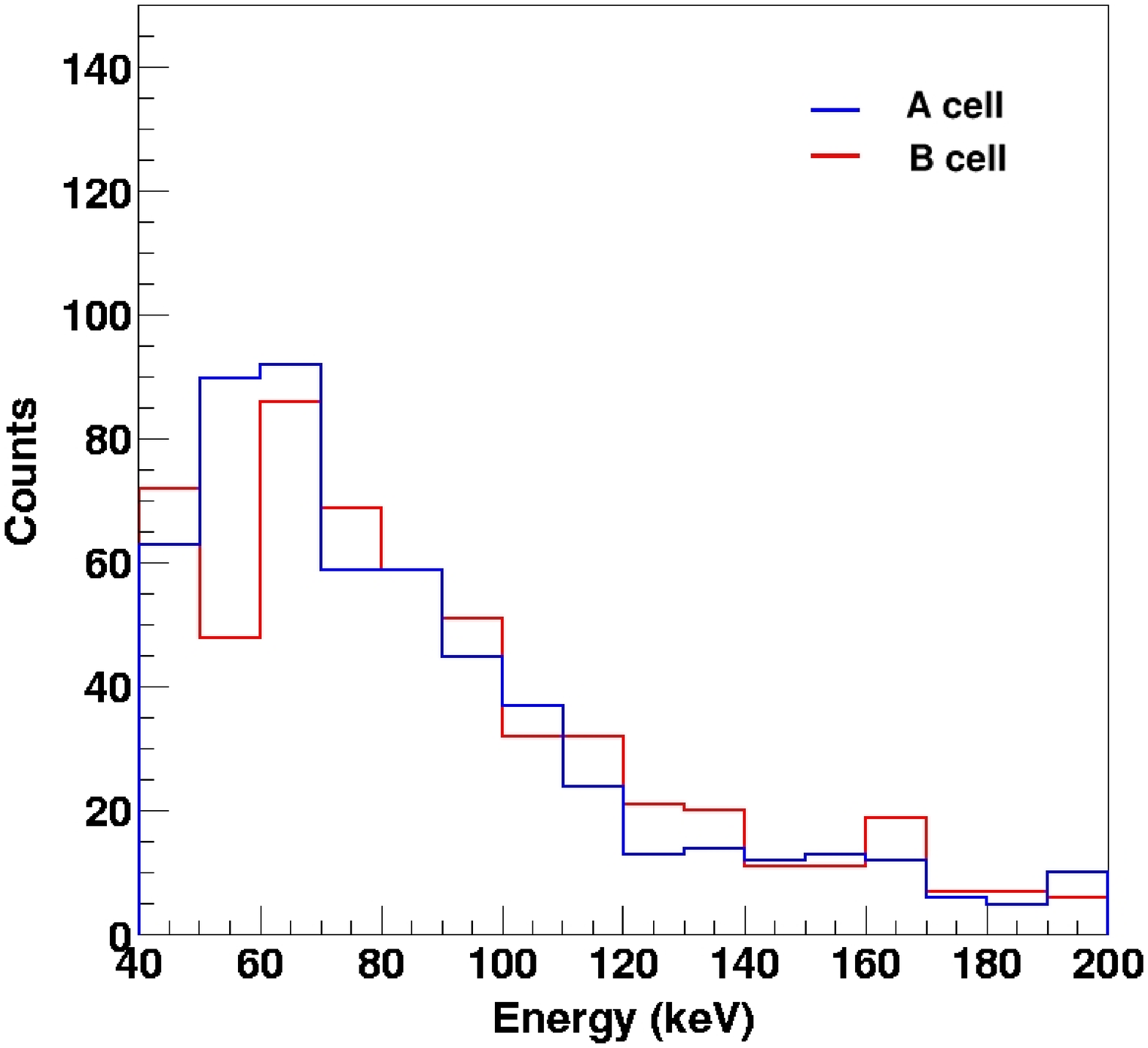,height=2.5in}
\caption{Acquisition data from cells A and B (a). Events in coincidence
are clearly visible. They are composed of cosmic muons. The deposited energy
by incident muons depends on the track length they makes inside the cell.
Cosmic muon detection in A and B cells (b). Spectra from cells A and
B exhibit a peak at 65 keV.\label{fig:muon}}
\end{center}
\end{figure}

Figure~\ref{fig:muon} (a) displays acquisition spectra from cells A and B.
Events in coincidence are clearly visible. The correlation among the cells
is proved to be a criteria for background rejection~\cite{nima}. The
coincidence rate~\cite{Clemens} among two cells is found to
be \rm{0.17 min$^{-1}$}.
The muon spectrum has been obtained during 18 hours around 100~\muKs.
It is presented on figure~\ref{fig:muon} (b).
The spectrum below 40 keV is shown in figure~\ref{fig:specelec}.
Spectra of two different cells agree very well. The muon peak is at about
{\rm 60 }keV. A more precise estimate of the poition of the muon peak and its comparison to theoretical expectations are  in progress ~\cite{Clemens}. The broadness of the peak is due to the 
fact
that the incoming muons release an amount of energy
related to the lenght of the tracks they produce inside the cell. Simulations
are in good agreement with the experimental with of the line.
\section{Conclusions}
A high sensitivity ultra-low temperature thermometry has been
developed as well as a working multicell prototype.
We demonstrated the possibility to detect low energy events by
embedding a low activity  electron conversion
source of \Co inside one cell. A spectrum presenting energy peaks at
7.3 and 13.6 keV
at 100 \muK is exhibited.
Furthermore we demonstrate the possibility
to improve background rejection by multicell correlation for a future
detector.

\section*{Acknowledgments}
This project was partially funded by the
R\'egion Rh\^one-Alpes, by the European
Community under the Competitive and Sustainable Growth programme (Contract
G6RD-CT-1999-00119), by the Bureau National de M\'etrologie (Contract 00 3
004). Part of this work was done in the framework of the European Science
Foundation scientific network "Coslab".
E. Moulin wishes to acknowledge financial support from a European
Union Grant for Young Scientists.

\end{document}